\documentclass[jcp,aps,groupedaddress,12pt]{revtex4}

\usepackage{amsmath}
\newcommand{\dd}{\,{\mathrm d}}
\newcommand{\vn}[1]{\mbox{\boldmath$#1$}}
\usepackage{graphicx}
\usepackage{afterpage}
\begin{document}

\title{A Concerted Variational Strategy for Investigating Rare Events}
\author{Daniele Passerone, Matteo Ceccarelli and Michele Parrinello}
\affiliation{CSCS - Centro Svizzero di Calcolo Scientifico, via Cantonale,
 CH-6928 Manno  \\
and \\
Physical Chemistry ETH, H\"onggerberg HCI, CH-8093 Zurich, Switzerland.}

\begin{abstract}
A strategy for finding transition paths connecting two stable basins is
presented.
The starting point is the Hamilton principle of stationary action; we show how
it can be transformed into a minimum principle through the
addition
of suitable constraints like energy conservation.
Methods for improving the quality of the paths are presented: for example, the Maupertuis
principle can be used for determining the transition time of the trajectory
and for coming closer to the desired dynamic path.
A saddle point algorithm (conjugate residual method) is shown to be efficient
for reaching a ``true'' solution of the original variational problem.
\end{abstract}
\maketitle
\clearpage
\newpage

\section{Overview}\label{sec:1}

In many of the complex systems that are encountered in physics, chemistry and biology, the potential energy surface 
exhibits deep minima separated by large barriers. In such cases the dynamics is characterized
by long periods in which the system stays in one minimum, followed by a jump of very short duration into another minimum.
These transitions from minimum to minimum are rare but crucial since they reflect important 
changes in the system, such as chemical reactions or conformational modifications of molecules in solution.
The time interval between these rare but important events can easily exceed the limit of molecular dynamics (MD) simulations,
since it depends exponentially on the barrier height. 
In the case of a smooth potential energy surface (PES) transition state theory \cite{tst} can be used and a variety of methods
have been devised to locate the PES saddle points \cite{dimer}. However,
in a complex system the PES is rough and exhibits a very large number of saddle points, thus calling into question the very 
notion of transition state \cite{tst,bolhuis}. 
Given the relevance of this problem, several ways of studying the PES in complex systems have been proposed 
\cite{flooding,voter1,voter2,voter3,puddle}. These methods are based on the idea of accelerating a traditional 
MD through modification of the PES in uninteresting regions, through a suitable rescaling of temperatures, or via an increase of atomic masses in hydrogen-rich systems \cite{berendsen}. 

A different approach has been proposed by Pratt \cite{pratt}. Here the idea is to sample 
the ensemble of paths that connect one minimum with another. This idea has been developed into a 
powerful scheme by Chandler and collaborators \cite{chandler}. The resulting transition path sampling (TPS) method has shown great promise in the study of 
rare events \cite{chand1,dellago,bolhuis}.

In this approach a dynamical path that connects two minima is determined. Starting from this initial path a sound statistical mechanics path sampling procedure is put into place that allows transition states to be determined.
The procedure for finding the initial path is, however, far from simple and a variety of tricks \cite{chand1,dellago} have been used to this end.

Motivated by these considerations, we develop here a procedure
that is capable of obtaining a realistic dynamical path once the initial and final positions are given.
This is at variance with other path-based methods which aim at locating the TS or finding the minimal energy path
\cite{jons}.
We expect two benefits from our approach: one is to find reaction mechanisms in an unbiased way, the other to 
extract appropriate reaction coordinates in a complex system.
The basis of our approach is the variational principles of classical mechanics.
In particular, the Hamilton principle states that every classical trajectory that starts from a configuration $\vn{q}_A$ and ends in $\vn{q}_B$ after a time $\tau$ renders
the action (which we call $S_H$) stationary.
From this principle, in fact, Lagrange equations of motion can be derived \cite{landau,goldstein}.
It has been shown that only for small $\tau$ or in the trivial case of a free particle is the stationary point a minimum \cite{librodilevi}.

This poses serious problems in numerical applications of the principle.
In fact, algorithms for locating saddle points are either computationally
very demanding or have a very small radius of convergence.
For instance, Cho, Doll and Freeman \cite{doll}, and 
Elber {\it et al.} \cite{elber,elbernew} have proposed a method based on the minimization of the squared norm of the gradient of the action.
However, this approach to saddle points is numerically challenging, since it amounts to worsening the condition number of the problem, and it
requires the evaluation of second order derivatives.

Another important variational principle that does not require {\it a priori} knowledge of $\tau$ is the Maupertuis principle. In this 
principle the energy is preassigned and an action is defined that is stationary relative to the geometrical trajectory. This is 
complemented by a relation between the geometrical trajectory and the time elapsed, which allows $\tau$ to be determined.

Our strategy for finding transition paths is the following.
We estimate a reasonable value of $\tau$ and start from the Hamilton principle. 
The saddle point problem is circumvented by adding a penalty function which imposes energy conservation.
The resulting trajectory is then used as an input for
the Maupertuis principle and a new and improved value of $\tau$ is extracted. With this new value of $\tau$ one can go back to the
Hamilton principle and repeat the procedure until convergence. 
The resulting trajectories are extremely accurate, so much so that one lands in a region where a quadratic expansion of $S_H$ is valid, and numerical methods can be applied to the
 unmodified Hamilton action in order to find the saddle point \cite{saad}.

This paper is organized as follows: in section \ref{sec:1} we introduce our method based on the Hamilton and Maupertuis principles; in sections
\ref{sec:2} and \ref{sec:3} we describe in detail the computational algorithms we use; in section \ref{sec:4} we present a working example: the application
to alanine dipeptide; finally, section 
\ref{sec:6} is devoted to our conclusions.

\section{Hamilton principle and Maupertuis principle}\label{sec:2}

In a previous paper \cite{noiprl}, we introduced an action-based optimization procedure,
relying on the well-known Hamilton principle. This principle can be stated as follows.
Let us consider a classical system with $3N$ degrees of freedom, characterized by its Lagrangean $L(\vn{q},\dot {\vn{q}})=T-V$, where $T$ and
$V$ are the kinetic and potential energy respectively. If the multidimensional Lagrangean coordinates $\vn{q}(t)$ are fixed to have the values $\vn{q}(0)=\vn{q}_A$ and
$\vn{q}(\tau)=\vn{q}_B$ at the initial and final time $0$ and $\tau$, then the dynamical path can be obtained by making stationary the action
\begin{equation}\label{eq:S}
S_H=\int_0^{\tau}L ( \vn{q},\dot {\vn{q}},t ) \dd t
\end{equation}
for all the variations that keep the extrema values $\vn{q}_A$ and $\vn{q}_B$ fixed. 
Indeed, Lagrange's equations of motion are derived directly from (\ref{eq:S}).
The stationarity does not guarantee that $S_H$ is either a minimum or a maximum. On the contrary: only for 
sufficiently low values of $\tau$ is the stationary point of $S_H$ a minimum \cite{librodilevi}, and Jacobi \cite{jacobi} has shown that it can never be a maximum; in fact the 
most common occurrence is a saddle point.
We now consider a dynamical path $\vn{q}(t)$ obtained by integration of Lagrange's equations. According
to Hamilton's principle the $S_H$ is stationary and we can consider the behavior of 
the Hessian $\delta^2 S$ as a function of $t$. As discussed above, for small $t$ the $S_H$ is a minimum and therefore the Hessian has all positive eigenvalues. As 
the system evolves this character is lost and the time at which the first zero eigenvalue of the Hessian sets in defines the conjugate point. It can be proved that, once $\vn{q}(t)$ has passed 
through this point, the trajectory is no longer a minimum; for larger $t$, new positive 
eigenvalues change sign. In the harmonic oscillator, for example, a new conjugate point occurs at each half period of oscillation.
Only if the system has a sufficiently small number of conjugate points will the number
 of negative eigenvalues be much smaller than the dimension of the space, and these directions could be individuated and cured, as shown in \cite{gillilan}. This is not the case for trajectories of interest, as 
we will show in detail in a forthcoming paper \cite{tobepub}.

If the system is conservative the total energy $E$ is a constant of motion. In such a case, Maupertuis has derived a variational principle in which the time does not appear explicitly (historically, the so-called least action principle of Maupertuis came before 
Hamilton's principle). If we write the geometrical trajectory in the parametric form $\vn{q}(s)$, where $\vn{q}(0)=\vn{q}_A$ and $\vn{q}(\sigma)=\vn{q}_B$, the
Maupertuis action for a system of $N$ particles, with $3N$ degrees of freedom of mass $m_i$, in the so-called Jacobi's form, reads:
\begin{equation}\label{eq:S0better}
S_M=\int_0^{\sigma} \dd s \sqrt {2(E-V)} \sqrt{\left(\sum^{3N}_{i=1} m_i \frac{\dd q_i}{\dd s} \frac {\dd q_i} {\dd s}\right)},
\end{equation}
and $S_M$ is made stationary. In (\ref{eq:S0better}) the total time $\tau$ can be calculated as
\begin{equation}\label{eq:time}
\tau=\int_0^\sigma \dd s \sqrt \frac {\sum^{3N}_{i=1} m_i \frac{\dd q_i}{\dd s} \frac {\dd q_i}{\dd s}}{2 (E-V)}.
\end{equation}

$S_M$ is plagued by the same problem as $S_H$ since its Hessian does not have a definite 
character; furthermore $(E-V)$ is not positive definite, which can cause problems in 
the search for a stationary state.
In spite of these difficulties, we will show later how to use this principle to obtain realistic paths.
\subsection{Inverted potential and second order methods}

The occurrence of conjugate points during the time evolution of a system represents an obstacle to the development of efficient optimization algorithms.
Almost 20 years ago Gillilan and Wilson  \cite{gillilan} (GW)
discussed how to overcome this problem in the search for dynamical paths.
First of all, GW introduced a discretized Hamilton principle (\ref{eq:S}), with the integral substituted by a sum, and the velocities defined by:

\begin{equation}\label{eq:vdiff}
\widetilde {\dot {\vn{q}}}(l)=\frac{\vn{q}^{(l+1)}-\vn{q}^{(l)}}{\Delta},
\end {equation}

so that the discretized action $\widetilde{S_H}$ reads

\begin{equation}\label{eq:discreteS}
\widetilde{S_H}:=\Delta \sum_{l=1}^{P-1} \left(\frac{1}{2} m  \left(\frac{\vn{q}^{(l+1)}-\vn{q}^{(l)}}{\Delta}\right)^2-V\left(\vn{q}^{(l)}\right)\right).
\end{equation}

Applying the 
condition of stationarity to the function (\ref{eq:discreteS}) one obtains a 
set of discrete equations of motion, corresponding to the well-known Verlet algorithm:

\begin{equation}\label{eq:verlet}
\vn{q}^{(l+1)}=2 \vn{q}^{(l)} -\vn{q}^{(l-1)}-\frac{\Delta^2}{m} V'\left(\vn{q}^{(l)}\right).
\end{equation}

GW pointed out that the problem defined by $\widetilde{S_H}$ is isomorphic to that
of a polymer on a surface. The points along the path are the beads of the polymer
and the kinetic energy term provides the
harmonic force that holds the polymer together.
Eq. (\ref{eq:verlet}) expresses the equilibrium between the harmonic forces and the forces coming from the surface. In this isomorphism  the saddle point nature of the
stationary point is reflected in the unstable equilibrium between the elastic force and the derivative of the potential $ V'\left(\vn{q}^{(l)}\right)$. Changing the sign of $V$ leads 
instead to a stable situation and  $\widetilde{S_H}$ has a minimum. By changing the sign of $V(\vn{q})$ one recovers the elastic band method.

Several methods  
for finding transition states and reaction paths \cite{elber,jons} have been developed on the basis of this property.
Continuing with the polymer analogy, these methods often fail to keep the beads evenly spaced along the path; in particular, a consequence of 
this fact is that the polymer does not pass exactly through the transition state, but overestimates the saddle point energy by cutting 
the PES (``corner cutting'') around the saddle point.

From a more formal point of view one can observe the correspondence between the 
inverted potential path and the instanton semiclassical theory. On this basis one can describe corner cutting as a manifestation of quantum tunneling \cite{wkb}. 
If one needs to locate the transition state, this effect is undesired and the nudged 
elastic band method has been devised to correct for it \cite{jons}.

A standard method to search for a stationary point is to look for the minimum of the sum of the squares of the derivatives. 
When applied to our case, the object function $S_{OM}$ to be minimized is:

\begin{equation}
S_{OM}:=\sum_{l,i}\left(\frac{\partial \widetilde S_H}{\partial q_i^{(l)}}\right)^2=\sum_l\left(q_i^{(l+1)}-2 q_i^{(l)} +q_i^{(l-1)}+\frac{\Delta^2}{m} V'\left(q_i^{(l)}\right)\right)^2,
\end{equation}
which becomes a minimum when $\widetilde S_H$ is stationary. Interestingly enough this action coincides with the discretized version of the Onsager-Machlup  action (OM-action) introduced in \cite{doll,elber} from a totally different point of view.
Minimizing this action requires the computation of the second derivatives of the potential energy, which can be rather demanding and even prohibitive for
ab-initio simulations. Furthermore, since the effective condition number of the problem is 
squared, obtaining accurate solutions is difficult and the resulting trajectories exhibit poor energy conservation \cite{noiprl}.
However, since $S_{OM}$ measures the deviation of the trajectory from a Verlet one, we shall use its value to assess the quality of a path.

\subsection{Our method}

A consequence of the Hamilton principle is energy conservation along the path (for
systems where the Hamiltonian does not depend explicitly on time).
In a discretized path that obeys (\ref{eq:verlet}) at each time step, energy conservation is no longer exact (and becomes worse and worse
as $\Delta$ is increased), but 
will fluctuate about an average value with a certain variance. In a first approximation, this effect can be produced by
adding to the original action a potential that keeps the energy oscillating around an equilibrium value with a given force constant.
We decided therefore to introduce and minimize the following modified action $S_\Theta$ \cite{noiprl}:

\begin{equation}\label{eq:ouraction}
S_\Theta=\gamma \int_0^{\tau}L(\vn{q},\dot {\vn{q}},t)\dd t  + \mu \int_0^{\tau}(T + V - E)^2 \dd t ,
\end{equation}

where only the ratio between $\mu$ and $\gamma$ matters, and $\gamma$ can be positive or negative. In the following we restrict the choice to the two values $\gamma=\pm 1$. The latter term ensures the conservation of energy.
In the non-discrete case, the second integral is exactly zero for physical paths, whereas it will be small and positive if the integral is discretized. Imposing energy conservation, although redundant, has the advantage of driving the minimization algorithm toward a 
reasonable zone of the path space in a natural manner.
The value of $E$ in (\ref{eq:ouraction}) can be also used as a parameter in the first stages of 
minimization, in order to guide the system toward a value of total energy compatible with the total time $\tau$.
Within this scheme it is also possible to implement constraints involving other conserved quantities like total linear momentum and angular momentum; in this paper we will refer solely to energy conservation.

Czerminsky and Elber \cite{czerm}
have already noted in passing that imposing energy conservation through a particular choice of parameters in their method could be useful but have
not subsequently used this remark.
Here instead energy conservation plays a major role and it is the 
crucial ingredient for arriving at correct dynamical trajectories in 
an efficient and numerically stable way.

From the mathematical point of view, the consequence of the second term in  (\ref{eq:ouraction}) is that the new functional has a minimum for positive values of $\mu > \mu^{\star}$. We have demonstrated  this property empirically in ref. \cite{noiprl} for simple cases. 
 Moreover, apart from the condition $\mu > \mu^{\star}$, the value of $\mu$ is not critical for the location of the solution; this behavior has also been observed in more complicated systems.

A solution obtained from the minimization of (\ref{eq:ouraction}) will in general differ from the solution of the Hamilton principle. In other words, one has to refer to quality factors such as the Onsager-Machlup action $S_{OM}$ in order to judge whether the solutions are dynamically sound.  

In the following, we will define:

\begin{itemize}
\item $S_H$-path as a solution of the original variational problem (\ref{eq:S}) in the continuum limit;
\item $V_{\Delta}$-path as a path obtained by integration of Verlet equations of motion, obtained from a discretized Hamilton principle. For a small enough $\Delta$, a $V_{\Delta}$-path can be considered a faithful sampling of a continuous $S_H$-path;
\item $\Theta$-path as a solution of minimization of the functional $S_\Theta$. 
\end{itemize}

The total time of the path can be estimated using physical or
chemical considerations.
However, this initial guess can be further refined by an alternate use of $S_\Theta$ and $S_M$. A similar, but more expensive approach (since it requires the second derivatives of the energy) has been applied recently by Elber {\it et al.} to the computation of trajectories \cite{elbernew} with very large time steps. Our procedure is instead as follows.

\begin{itemize}
\item First, we fix points $\vn{q}_A$ and $\vn{q}_B$ in configuration space and calculate $V(\vn{q}_A)$ and $V(\vn{q}_B)$. They do not need to be two exact minima of the potential: two points in the basin of attraction of the two states $A$ and $B$ are sufficient in this context. A good procedure is to sample these points from traditional MD simulations performed in the basins of $A$ and $B$.
\item We need an initial guess for the path. To start from the linear path connecting $\vn{q}_A$ and $\vn{q}_B$ is the simplest choice. 
From this linear interpolation, we can easily obtain a rough approximation of a minimum energy path either by inverting the sign of $V$ in (\ref{eq:discreteS}) and minimizing that functional, or by minimizing $S_\Theta$ with $\tau$ set to a computationally convenient value, and a conserved energy $E_{low} < V_A$. The minimum energy path (although not dynamical) gives a crude estimate of the barrier. This allows us to set a first choice for $E$, the total energy along the path, slightly above the barrier.
\item A first value for the total time $\tau$ can then be obtained from  (\ref{eq:time}) applied to the minimum energy path
treated as a geometric trajectory. Through this equation, we obtain a non-uniform time distribution of the intervals along the path.
\item We map the set of geometric intervals on this $S_M$ trajectory onto a set of time intervals 
on the $\Theta$-path, and we redistribute the points of the path on an even grid.
\item With the $\tau$ obtained at the previous iteration, we use $S_\Theta$  full minimization (possibly preceded by a slight randomization of the previous path) and find a local minimum.
\item This solution is used as starting point for a few steps of $S_M$ partial minimization. The path will be modified and $\tau$ will be slightly corrected.
Partial minimization can be successfully replaced by a few cycles of the conjugate residual algorithm for finding undefined stationary points described later in this paper. 
\item The procedure is repeated until a lower threshold for a quality factor of the path (such as $S_{OM}$) is reached.
During the iterative scheme the value of $E$ can be adjusted automatically, treating it as a slow degree of freedom during the
minimization of $S_\Theta$. The value of $\tau$ will change accordingly during the $S_M$ phase of the optimization.
\end{itemize}

The quality of the paths is very high and for many purposes the calculation can be stopped here. However, we shall show in the following that the quality can be further improved and that the exact variational solution of the discretized Hamilton's principle can be obtained, that is, a $V_{\Delta}$-path can be calculated.

\subsection{Fourier components and integration}

The paths can be described numerically in various ways. The simplest one is to use the discretization used in eq. (\ref{eq:discreteS}). In most of the applications we have
instead used a Fourier expansion. Following Cho {\it et al.}\cite{doll}, we write the trajectory as:

\begin{equation}\label{eq:rfou}
\vn {q} (t)= \vn {q}_A + \frac{(\vn {q}_B - \vn {q}_A)t}{\tau}+\sum_{n=1}^{P-1} \vn {a}^{(n)} \sin (\frac{n \pi t}{\tau}).
\end{equation}
In such a way we implicitly take into account the boundary condition 
$\vn{q}(0)=\vn{q}_A$ and $\vn{q}(\tau)=\vn{q}_B$.
The advantage of this choice is that one can start with few components, thus
obtaining smooth trajectories which represent a good initial guess and
avoiding local spikes in the coordinates which lead to unreasonable values of the kinetic energy.
Since in this representation the positions are defined at any $t$, the velocities can be exactly calculated as:

\begin{equation}
\widetilde {\dot {\vn {q}}}^{(l)}=\frac {\vn {q}_B-\vn{q}_A}{\tau} + \sum^P_{n=1} \frac{\pi n}{\tau}\, \vn {a}^{(n)} \cos \frac{\pi l \Delta n}{\tau}.
\end{equation}

With this choice, even if the path is discretized, particle velocities are defined at every point of the trajectory.
Although (\ref{eq:rfou}) defines the trajectories in a continuum fashion, the numerical evaluation of the integral that defines $\widetilde{S_\Theta}$ requires 
the use of a discrete mesh. 
In principle the action integral does not need to be evaluated on a $P$ point
mesh, but this seems the natural choice, since it is compatible with the path representation.
In this way we have only one convergence parameter, and the larger the value of $P$, the better the path description. 
With this choice one has:

\begin{equation}\label{eq:tildeth}
\widetilde S_\Theta = \sum^P_{i=0} w_i \Delta \left( \gamma L(\{\vn{q}^{(i)}\},\{\dot {\vn {q}}^{(i)}\})+\mu (T^{(i)}+V^{(i)}-E)^2 \right),
\end{equation}
where $\Delta=\frac{\tau}{P}$ and the weights $w_i$ depend on the integration 
algorithm.
For instance, using Simpson's rule, which will be our default choice, $w_0=1/2$, $w_i=1$ for $i=1\dots(P-1)$, and $w_P=1/2$.

\section{Minimization and optimization algorithms. Refining the trajectory}\label{sec:3}

Since $\widetilde S_{\Theta}$ can have several minima, in order to minimize our $S_\Theta$ action we use not only a conjugate gradient (CG) algorithm, but also a simulated annealing (SA) 
procedure.
In fact CG has the disadvantage of leading toward a nearest minimum; with a simulated annealing MD, we assign a fictitious mass to the Fourier components
${\vn {a}^{(n)}}$ and a ``temperature'' to the system, first evolving it at high temperature in order to better explore the parameter space, and
then ``cooling''  it toward a minimum of the $S_\Theta$.
This minimum can subsequently be refined through the CG algorithm.

In any case, during the algorithm we have to calculate the derivatives of the action  $S_\Theta$ with respect to the Fourier components and set them to zero. We write here 
the explicit form for these equations:

\begin{eqnarray}\label{eq:gradient}
\frac{\partial S_\Theta}{\partial a^{(n)}_{i}}&  =  & 
\Delta \sum_l w_l m_i \nonumber \\
(\frac {\vn {q}_B-\vn{q}_A}{\tau}  +  
\sum^P_{m=1} \frac{\pi m}{\tau} &\vn {a}^{(m)}& \cos \frac{\pi l \Delta m}{\tau} ) \frac{\pi n}{\tau}\, \cos \frac{\pi l \Delta n}{\tau} 
\left(  1+ 2 \mu \left(T^{(l)}+V^{(l)}-E \right) \right)
+ \nonumber \\
F_{i}^{(l)}  (\{\vn {a}^{(n)}\}) &\sin& \frac{\pi l \Delta n}{\tau}
\left(  1- 2 \mu \left(T^{(l)}+V^{(l)}-E\right) \right)=0 \nonumber \\
 \,\,\, (i=1\dots 3N,n=1 \dots P-1),
\end{eqnarray}
where $F_i^{(l)}$ is the $i$-th component of the force on the ions at time slice
$l$.

\subsection{Specific implementations of the algorithm}

We have developed a computer program for performing the minimization of the functionals
described in this paper. 
The program {\bf VERGILIUS} is an interface connecting the evolution of the path to the calculation, at every time slice, and for every step of the optimization, of the potential energies ${V}$ and of the forces $\vn{F}$ requested by equations (\ref{eq:gradient}).

At present, {\bf VERGILIUS} is interfaced to the plane wave MD code {\bf CPMD}, a quantum chemistry code (GAUSSIAN 98 \cite{gaussian}), and to the biomolecular simulation package  ORAC\cite{orac}.

The algorithm is naturally parallel, but since we adopt a Fourier representation, there are two additional communication phases among 
the nodes of a parallel machine performing a $S_\Theta$-minimization: the distribution of the coordinates from the Fourier components and the building of the gradient. 
To make this phase efficient, we have adopted a fast Fourier transform (FFT) algorithm \cite{fftw}, and the scaling of the overall computational scheme is very good.
 
\subsection{Quality factors for the solution. Fourier norm}

Once a trajectory is obtained, the question of its accuracy arises.
As already stated, if the time step is sufficiently small, the closeness of the trajectory to a Verlet trajectory is 
a good criterion: the lower the value of the OM-action, the better the trajectory. In this section we will briefly describe this criterion, possible problems arising from it, and some procedures for judging the physical soundness of the solution found. 

Since we are optimizing or minimizing our action in Fourier space, another functional introduced by Cho, Doll and Freeman \cite{doll} can 
be more appropriate, namely the norm of the gradient of $S_H$ (that is, (\ref{eq:gradient}) with $\mu=0$) in Fourier space (F-norm):

\begin{equation}
\sum_{i,n}\left(\frac{\partial S_{\Theta}}{\partial a_i^{(n)}}\right)_{\mu=0}=\sum_{i,n} (\Delta \sum_l w_l m_i \widetilde {\dot q_{i}^{(l)}}  \frac{\pi n}{\tau}\, \cos \frac{\pi l \Delta n}{\tau}+
F_{i}^{(l)} \sin \frac{\pi l \Delta n}{\tau})^2.
\end{equation}

Here $i$ is the particle component index (ranging from 1 to $3N$) and $l$ is the time slice index (ranging from 1 to $P-1$).
Due to the definition of the velocities, this norm is not exactly zero for a $V_\Delta$-trajectory. We have verified, however, that there is a one-to-one correspondence between F-norm and OM-action, such that a low value of the F-norm always corresponds to a trajectory that is close to a $V_{\Delta}$ path.

\subsection{Refining: how to go from a $\Theta$-path to a $V_{\Delta}$-path. Conjugate residual method}

Once a $\Theta$-path is obtained (possibly improved through the $S_\Theta$-$S_M$ iterative scheme) and a reasonably low value for the OM-action found, we can try to reach a stationary point of the original variational problem. Having kept the discretization time step sufficiently small, that is, comparable with the fastest intrinsic vibrations of the system, we can be confident that the $V_{\Delta}$-path we are aiming at is a good representation of a $S_H$-path.

For finding the saddle point of $S_H$, we use an algorithm described in ref. \cite{saad}, namely the Conjugate Residual (CR) method, which is very similar in spirit to the CG, but is not limited to solving positive definite linear systems.
This algorithm is most efficient for sparse systems. The 
choice of a suitable parameter space is therefore crucial; whereas for a $S_\Theta$ minimization a 
Fourier component representation has many advantages, 
we found that once the optimization procedure is sufficiently close to a $V_{\Delta}$-path, returning to Cartesian coordinates with velocities defined as finite differences can be wise, since the Hessian of  (\ref{eq:ouraction}) is naturally sparse in this representation.  
We therefore adopt for this refining phase the representation in Cartesian components $\{\vn{q}\}$.

We assume that we are close to the stationary point and expand the action $S_H$ around the position $\{\vn{q}^{0}\}$:

\begin{equation}
\tilde S_H (\vn{q}) \simeq \tilde S_H (\vn{q}^0) + \sum_j (q_j - q^0_j) \left.\frac{\partial \tilde S_H}{\partial q_j}\right|_{\vn{q}^0}+\frac{1}{2}\sum_{h,k} (q_h-q_h^0)\left.\frac{\partial^2 \tilde S_H}{\partial q_h \partial q_k}\right|_{\vn{q}^0} (q_k-q_k^0).
\end{equation}
The condition of zero gradient leads to the linear system
\begin{equation}
\hat A \vn{x} = \vn {b},
\end{equation}
where
\begin{equation}
A_{ij}:=\left.\frac{\partial^2 \tilde S_H}{\partial q_i \partial q_j}\right|_{\vn{q}^0}\;;\;b_j:=-\left.\frac{\partial \tilde S_H}{\partial q_j}\right|_{\vn{q}^0}\;;\;\vn{x}:=\vn{q}-\vn{q}^0.
\end {equation}

The algorithm scheme is, then, in pseudo-code \cite{saad}: \\

\begin{ttfamily}
 Compute $r_0:=b-Ax_0$,$p_0:=r_0$

 For $j=0,1,\dots, $ until convergence Do: 

 $\;\;$ $\alpha_j := (r_j,A r_j)/(A p_j,A p_j)$ 

 $\;\;$ $x_{j+1}:=x_j + \alpha_j p_j$

 $\;\;$ $r_{j+1}:=r_j - \alpha_j A p_j$

 $\;\;$ $\beta_j := (r_{j+1},A r_{j+1})/(r_j,A r_j)$ 

 $\;\;$ $p_{j+1}:=r_{j+1}+\beta_j p_j$

 $\;\;$ Compute $A p_{j+1} = A r_{j+1} + \beta_j A p_j$

 EndDo \\
\end{ttfamily}

This algorithm is the counterpart of the conjugate gradient for non-positive definite matrices.
The only matrix-vector product to be performed is the product $A r$.
The Hessian of $S_H$ (with the velocities defined as in ($\ref{eq:vdiff}$)) is a matrix of linear dimension $(P-1) \times 3N$ (with $P$ the number of time slices, and $N$ the number of particles), and has the following block diagonal form:
\fontsize{8}{1}
\begin {multline}
A_{(ti,fi),(tj,fj)}=\frac{\partial^2 S_H}{\partial q_{fi}^{(ti)} \partial q_{fj}^{(tj)}}= \\
 \begin{pmatrix}
\Delta   (w_{t1} H_{t1} + w_{t1} \frac {\vec m }{ \Delta^2 }I) & -\Delta w_{t1}
\frac {\vec m}{\Delta^2} I  & 0 & 0 & 0 & 0 \\
-\Delta w_{t1}\frac {\vec m}{\Delta^2} I  &\Delta   (w_{t2} H_{t2} +(w_{t1}+w_{t2}) \frac {\vec m }{ \Delta^2  } I )& -\Delta w_{t2}\frac{\vec m}{\Delta^2}I  &
0 & 0 & 0\\
0& \hdotsfor {4}&0 \\
0& \hdotsfor {4}&0 \\
0 & 0 & \dots & 0  &  -\Delta w_{tP-1} \frac {\vec m}{\Delta^2} I  & \Delta   (w_{tP} H_{tP} + (w_{tP-1}+w_{tP}) \frac {\vec m }{ \Delta^2}I). \\ \nonumber
\end{pmatrix}
\end{multline}
\normalsize

Here $ti,tj$ are time slice indexes, $fi,fj$ are Cartesian components indexes, $H_{ti}$ is the Hessian of the potential at the time slice $ti$ and
a $3 N \times 3 N$ matrix; $\vec m$ is the vector of the masses and $w_i$ are the weights for discrete integration.
Despite the sparse nature of this matrix, the computation of the matrix blocks is demanding.
We chose therefore to
calculate this product using a first order expansion:

\begin{equation}\label{eq:trick}
(\hat A  \cdot \vn{q})_i = \sum_j \left.\frac{\partial^2 S_H}{\partial q_i \partial q_j}\right|_{\vn {q}} r_j \simeq \frac{1}{2 \lambda} \left( \left.\frac {\partial S_H}{\partial q_i}\right|_{\vn {q} + \lambda \vn {q}}- \left.\frac {\partial S_H}{\partial q_i}\right|_{\vn {q} - \lambda \vn {q}}\right),
\end{equation}
with $\lambda$ small. With this approximation, the algorithm remains sufficiently efficient if the solution is not too far from the starting point. 
Once convergence is reached, a new quadratic expansion of $S_H$ can be performed and the procedure is iterated until a desired value of the OM-action is found.

A good preconditioning of $A$ is desirable in order to accelerate convergence.
We found a good compromise between computational cost and efficiency using 
so-called ``diagonal preconditioning'', which amounts to modifying the original system $\hat A \vn{x} = \vn{b}$ into the system $\hat M^{-1} \vn{x} = \hat {M}^{-1} \hat A \vn{b}$, with the aim of reducing the condition number $\lambda_{\mathrm{max}}/\lambda_{\mathrm{min}}$ of the matrix $\hat A$, and $\hat M$ defined as

\begin{equation}
M_{hk}:=\left\{ \begin{array}{cc} A_{hk} & h=k \\ 0 & h \neq k\end{array}\right..
\end{equation}

A convenient way to parallelize the calculation is for instance to assign a time slice to each node.
Then the diagonal terms of the Hessian can be calculated exactly using eq. ($\ref{eq:trick}$) by substituting $\vn{q}$ with a set of $3N$ vectors $\vn {q}^{(l)}$ of dimension $3N \times P-1$, with zeros everywhere but in $P-1$ positions $l+3N (k-1), k=1 \dots P-1$.
For example, vector $\vn {q}^{(3)}$ has the form:
\begin{equation}
\begin{pmatrix}
0&0&1&(3N-3) \, \rm {zeros} &0&0&1& (3N-3) \, \rm {zeros} &\dots ,\\
&&&&&&&
\end{pmatrix}
\end{equation}
repeated $P-1$ times.
With this choice every parallel node will return the desired information about the diagonal part of $\hat A$: the application of eq. \ref{eq:trick}  to the set of vectors $\vn {q}^{(l)}$ leads to a set of vectors $G^{(l)}$, from which the known off-diagonal term due to kinetic energy must be subtracted. At this point the desired matrix $M_{hk}$ is formed, and the preconditioned algorithm can be applied. 

\section{A working example: alanine dipeptide}\label{sec:4}

The test system we will use in this section is a model of alanine dipeptide, a small peptide made of 22 atoms. 
The PES of this molecule in vacuum shows two main minima: an extended configuration, and an axial configuration \cite{bolhuis}. The barrier in vacuum is about 7 kcal/mol.
The conformational change from an equatorial state to an axial state has been studied thoroughly \cite{bolhuis,apostolakis}, in particular with a view to estimating free energies both in the vacuum and in solution.
As an example, we will adopt here the united atoms (UA) scheme. In this model, only two hydrogens are treated explicitly, whereas the others are 
contracted into superparticles with proper mass and charges, according to the OPLS-AMBER scheme \cite{opls,amber}.
The resulting UA molecule is made of 12 atoms. We thus have (36-5) independent degrees of freedom.

The scope of this section is to show how a $\Theta$-path is obtained and how the $S_M$ can be used for refining the path. The quality of the trajectory obtained is confirmed by the fact that CR algorithm easily reaches a $V_{\Delta}$-path using the refined $\Theta$-path as a trial trajectory.

In order to obtain a realistic path, we used exactly the iterative scheme described in Section \ref{sec:2}.
We started from a linear interpolation between the two minima. We alternated $S_\Theta$-action minimization (with $\gamma=-1$ and $\mu=25000$) and partial $S_M$-action minimization (using both simulated annealing and conjugate gradient algorithms) in order to obtain a dynamical path passing through the barrier.

\begin{figure}
\includegraphics[width=0.5\textwidth]{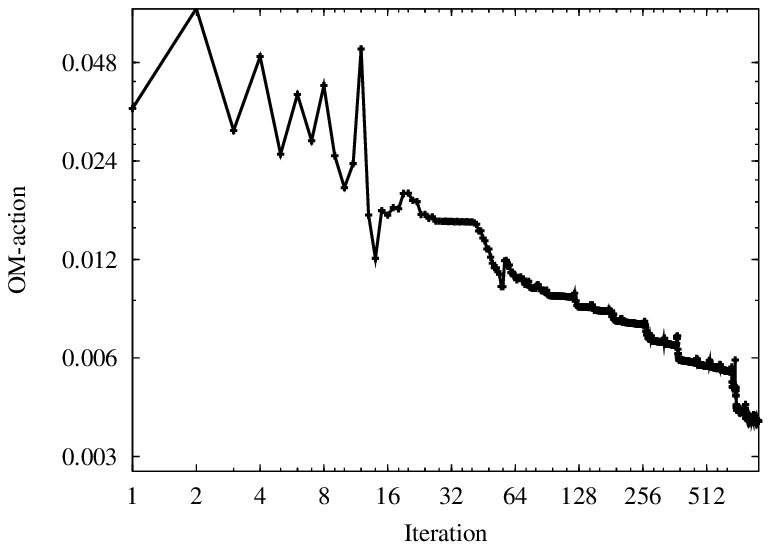}
\caption{Alanine dipeptide: value of the OM-action during the iteration of
the $S_\Theta$-$S_M$ optimization scheme. Even iterations correspond to 
$S_\Theta$ minimization, odd iterations to partial $S_M$ minimization.
}\label{fig:mau1}
\end{figure}

The results are shown in Fig. \ref{fig:mau1}, where 
we plot the value of the OM-action as a function of the iteration steps.
The $S_M$-action refinement allows the value of the quality factor to be reduced by one order of 
magnitude.

\begin{figure}
\includegraphics[width=0.99\textwidth]{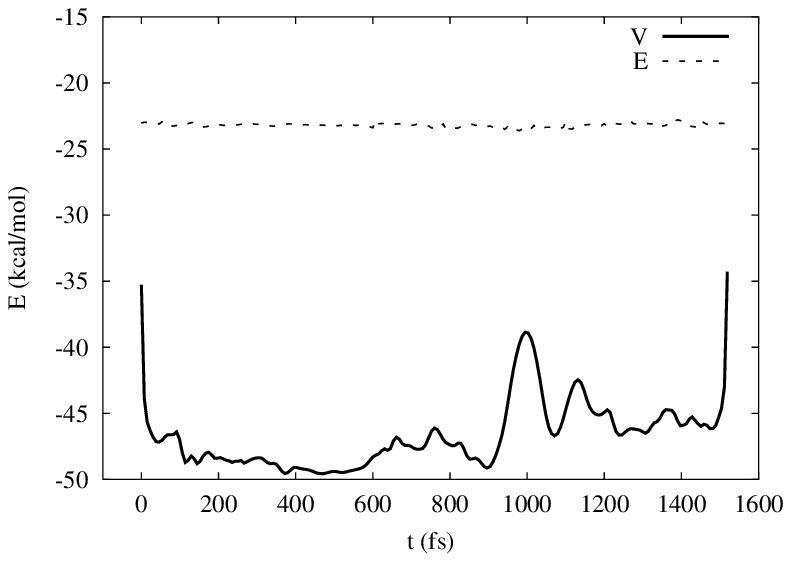}
\caption{Alanine dipeptide: potential energy V and total energy E profile for a path resulting from the $S_\Theta$-$S_M$ iterative 
procedure (refined) described in the text.
The total time has been determined by the algorithm, and is
qualitatively close to that set on the basis of phenomenological considerations about this rare event. The number of slices $P$ during the iterative procedure was fixed at 200.}
\label{fig:mhe}
\end{figure}

\afterpage{\clearpage}
The potential and total energy profiles for this Maupertuis-improved $\Theta$-path are shown in Fig. \ref{fig:mhe}.
The total time has been reduced from $\tau=2 \, \mathrm{ps}$ to $\tau=1.59 \, \mathrm{ps}$.

\subsection{Refinement} 
At this point, we are confident of being sufficiently close to a $S_H$-path to apply the CR method and find the stationary path of
Hamilton action.

We take the solution of the combined $S_\Theta$-$S_M$ strategy (a $\Theta$-path with $\Delta=1.9$ fs and $P=800$)  and 
use it as an input for our CR algorithm.

The algorithm allows a $V_{\Delta}$-path to be obtained with a value for the OM-action as low as $0.5 \times 10^{-6}$, $10^4$ times smaller than the initial value for the $\Theta$-path.
The resulting solution can be considered as a true dynamical trajectory, since $\Delta$ is small
with respect to the fastest oscillation period in this system.
The $\Theta$-path and the ``true'' path will diverge increasingly in the minimum region of the PES, where the  
trajectories are chaotic. 
This reflects a fundamental property of reactive trajectories, which exhibit chaotic behavior in the stable basins and regularities in the transition region if the energy gap above the saddle is not too large \cite{giappo}.
In terms of our algorithm the chaotic behavior can be explained as follows: near a potential energy minimum the 
forces (and the accelerations) are very small, and a small error in the acceleration can
cause a change of sign, leading to an incorrect curvature and a dramatic deviation of 
the approximate trajectory from the true one.

\begin{figure}
\includegraphics[width=0.6\textwidth]{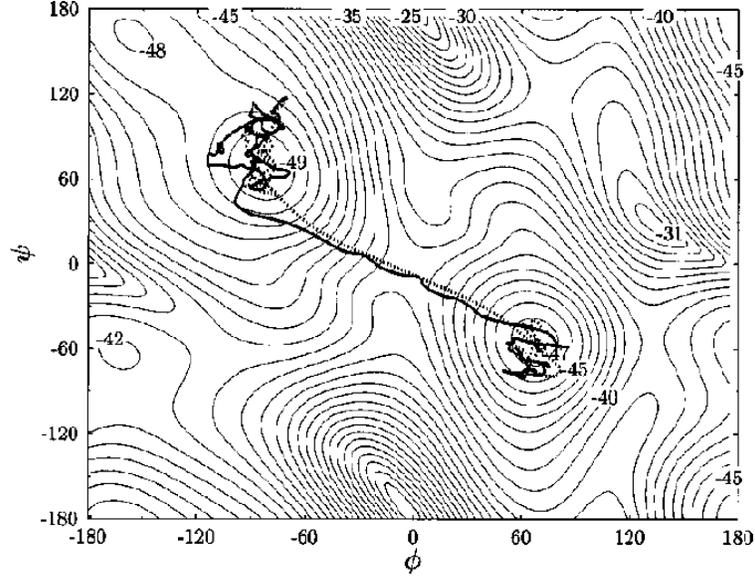}
\caption{Alanine dipeptide: $\phi-\psi$ Ramachandran plot of the $S_\Theta$-$S_M$ path (dotted line) and of the $V_{\Delta}$-trajectory (continuous line) obtained through the stationary point CR algorithm (see text). The two trajectories are very similar, apart from obvious differences especially in the minima regions, where the chaotic behavior of the
dynamics is more pronounced. 
}\label{fig:refining}
\end{figure}

To give a visual measure of the difference between the two trajectories, we plot
them on a $(\phi-\psi)$ PES, where $\psi$ and $\phi$ are the two soft dihedral angles of this molecule, as defined in  ref.
\cite{bolhuis}.
It can be seen in Fig. \ref{fig:refining} that the two paths pass through the
same transition state, and are also very similar in the two basins.

The Euclidean distance between the two trajectories $\{\vn{r1}\}$ and $\{\vn{r2}\}$, defined as

\begin{equation}
D=\frac{1}{(P-1) N} \sum_{l=1}^{P-1} \sqrt{\sum_{i=1}^{3N} (r1_i^{(l)}-r2_i^{(l)})^2}, 
\end{equation}
is in this case equal to $1.15 \, \mathrm{\AA}$/atom/slice.

Let us now look for a $V_{\Delta}$-path with $\Delta$ much smaller than that used up to now. Such a path can be useful, since in traditional MD the instability 
threshold for an integration algorithm (like Verlet) is well below that for a variational algorithm with two boundaries. Where a single transition state is 
present along the path we can proceed as follows.  We consider our $V_{\Delta}$ path obtained with the CR, and a small portion around the transition state.
Through interpolation, this small portion of the path is refined and $\Delta$ is reduced; CR is then used to find the stationary point of this short trajectory. 
At this point we consider two adjacent points in the discrete path. Forward and backward integration using the Verlet algorithm, with these points as initial conditions, lead to 
a reactive trajectory with a very small time step.
 The energy conservation will be of the same quality as in the original $\Theta$-path.

Concerning the choice of the parameters in (\ref{eq:ouraction}), we observed that the choice of $\gamma=-1$ leads to more realistic trajectories. We will explain this
fact in a forthcoming paper \cite{tobepub}.
The value of $\mu$ must instead be tuned in order to render the OM-action as small as possible.
Since there is a wide range of values of $\mu$ (several orders of magnitude) where this quality factor does not change
appreciably, tuning is not difficult.

\section{Conclusions and perspectives}\label{sec:6}

In conclusion, we have built a complete strategy for obtaining reaction paths 
close to solutions of the variational problem of classical mechanics.
We have presented an iterative procedure based on Maupertuis and Hamilton's principle; as a result, paths are found that are sufficiently close to dynamical trajectories to allow  a local projection algorithm like the conjugate residual method to be applied for  extracting the ``correct'' trajectory from the $\Theta$-path; this procedure  is promising also for larger systems.

The next step in our research will be to pass to phase space, where 
canonical momenta and coordinates have to be considered as independent variables.
Discretization of the Hamilton principle in phase space and application of
Euler equations to the resulting functional leads to the velocity Verlet 
algorithm of molecular dynamics.
Moreover, although the number of variables is double because of the
introduction of canonical momenta, the kinematic conditions (first 
equation of Hamilton) connecting the momenta to the particle velocities 
can be fruitfully used as further quadratic constraints, leading to 
another interesting optimization problem.

We wish to thank D. Aktah, A. C. Levi, A. Gusev and A. Laio for fruitful discussions.

\clearpage

\newpage

\clearpage

\end{document}